\documentclass[12pt,cite,epsf,epsfig]{article}
\usepackage{epsfig}
\title{Study of the Flavour Changing Neutral Current Decay 
mode $B\rightarrow K \gamma \gamma $}
\author{S.~R.~Choudhury$^a$\thanks{E--mail : src@ducos.ernet.in},
G.~C.~Joshi$^b$\thanks{E--mail: joshi@tauon.ph.unimelb.edu.au},  
Namit Mahajan$^a$\thanks{E--mail : nm@ducos.ernet.in, 
nmahajan@physics.du.ac.in} and
B.~H~.~J.~McKellar$^b$\thanks{E--mail:b.mckellar@physics.unimelb.edu.au}\\
	$a.$~{\em Department of Physics and Astrophysics,} \\
	 {\em University of Delhi, Delhi-110 007, India.}\\
	$b.$~{\em School of Physics, University of Melbourne,}\\
	{\em   Australia.}}

\def\be{\begin{equation}}
\def\ee{\end{equation}}
\def\bea{\begin{eqnarray}}
\def\eea{\end{eqnarray}}

\begin{document}
\maketitle
\large

\begin{abstract}
We study the neutral current flavour changing rare decay mode
$B\rightarrow K \gamma \gamma $ within the framework of
Standard Model, including the long distance contributions. It is
found that these long distance contributions can be larger
than the short distance contribution.\\ 
{\bf Keywords}:Rare B decay \\
{\bf PACS}:13.25.Hw, 13.40.Hq 
\end{abstract} 
------------------------------------------------------------------------------------------------------
We have corrected a sign error in one of the terms contributing to the
Irreducible (short-distance) contribution in our program. The revised
estimates agree well with the ones given in a recent paper \cite{hiller}.\\
------------------------------------------------------------------------------------------------------
\begin{section}{Introduction}
Radiative decays of the B-meson are very useful indices for testing
the underlying theories of flavour changing neutral currents (FCNC)
since the amplitudes are very sensitive to QCD contributions as well as
to possible contributions from loops with supersymmetric partner
particles. Of the radiative modes, the decay mode  $B\rightarrow X_s
\gamma$ has been measured experimentally and has also been the subject
of extensive theoretical investigations. The advent of the B-factories
will probably make possible  measurement of radiative decay modes with
two photons.  Hence the basic two photon amplitude $b\rightarrow
s\gamma\gamma$ has also been studied  starting with the work of Lin,
Liu and Yao \cite{lin}, and pursued further in references \cite{chang},
\cite{reina}.

The $b\rightarrow s \gamma \gamma $ amplitude falls naturally into two
categories; an irreducible contribution and a reducible one where the
second photon is attached to the external quark lines of the $b
\rightarrow s \gamma$ amplitude.  The irreducible part can be
estimated through the basic triangle graph.  Evaluating the reducible
part at the quark level then presents no problem.  For the process $B
\rightarrow X_s \gamma \gamma $ it is appropriate to the consider the
amplitude as dominated by the quark level amplitude.  However, when we
consider an exclusive channel $ B \rightarrow M \gamma \gamma $ with a
specific meson M, it is more appropriate to consider the reducible
contributions as arising out of emission of the second photon from the
external hadron legs of the amplitude $ B \rightarrow M \gamma $
rather than restricting oneself to quark level considerations and
thereby completly neglecting the spectator contributions  in
B and M. The amplitude $ B \rightarrow K
\gamma \gamma $ then stands out since the basic one photon amplitude $B
\rightarrow K \gamma $ vanishes for real photons and we may expect
the irreducible part of the amplitude to dominate.  There will of
course be the usual long distance contributions, the most important
being the process $ B \rightarrow K^{\star} \gamma $ followed by decay
of $K^{\star}$, and $ B \rightarrow K \eta_c $ followed by two photon
decay of the $\eta_c$.  The corresponding $\eta$ 
contribution will be small owing to the small $\eta$
coupling to $\bar{c}c$ but we include it also for completeness.
The $B \rightarrow K\eta^{\prime}$ amplitude however is anomalously
high, probably because of the gluon fusion mechanism. Since
$\eta^{\prime}$ has a sizeable branching ratio into $2~\gamma$, the
$\eta^{\prime}$ contribution to $ B \rightarrow K\gamma \gamma$ will 
prove to be 
the largest resonance contribution.  


This note reports an estimate of the branching ratio for the
process $ B \rightarrow K \gamma \gamma $ taking account of the irreducible
contribution as well as all significant resonance contributions. The 
earliest
investigations of this process restricted themselves to the $\eta_c$
contribution \cite{ahmady1}; subsequently an estimate
was made of a possible contribution from the
resonance $\eta^{\prime}$ with the $\eta^ {\prime}$ arising through gluon
fusion from the basic process $ b \rightarrow s g g $ and subsequently
decaying into two photons \cite{ahmady2}. The $ \eta^{\prime}$ resonance 
however is
very narrow and its contribution will show up as a distinct narrow
peak in the $\gamma \gamma$ invariant mass spectrum. This can experimentally
be easily separated and  we could restrict our analysis by excluding this
resonance, but we include it for completeness.   

There are of course interference terms from each pair of terms.  
Unfortunately the only interference term whose sign we can determine 
is that between the irreducible term and the $\eta_{c}$ resonance 
contribution.  The other signs are unknown.  We quote the largest and 
the smallest branching ratio, and give enough information for the 
reader to construct the other possibilities.

\end{section}

\begin{section}{Irreducible triangle contributions}
These arise from triangle diagrams with the photon being emitted by
the
 quark loop. The effective Hamiltonian is \cite{lin}
\be
{\cal{H}}_{eff} = -\frac{G_F}{\sqrt{2}}V_{ts}^*V_{tb}\sum_i 
                         C_i(\mu)O_i(\mu),
\ee
with
\[
O_1 = (\bar{s}_ic_j)_{V-A}(\bar{c}_jb_i)_{V-A},
\]
\[
O_2 = (\bar{s}_ic_i)_{V-A}(\bar{c}_jb_j)_{V-A},
\]
\[
O_3 = (\bar{s}_ib_i)_{V-A}\sum_q(\bar{q}_jq_j)_{V-A},
\]
\[
O_4 = (\bar{s}_ib_j)_{V-A}\sum_q(\bar{q}_jq_i)_{V-A},
\]
\be
O_5 = (\bar{s}_ib_i)_{V-A}\sum_q(\bar{q}_jq_j)_{V+A},
\ee
\[
O_6 = (\bar{s}_ib_j)_{V-A}\sum_q(\bar{q}_jq_i)_{V+A},
\]
\[
O_7 = \frac{e}{16\pi^2}\bar{s}_i\sigma^{\mu\nu}(m_sP_L +
m_bP_R)b_iF_{\mu\nu},
\]
and
\[
O_8 = \frac{g}{16\pi^2}\bar{s}_i\sigma^{\mu\nu}(m_sP_L + m_bP_R)
           T_{ij}^ab_jG_{\mu\nu}^a.
\]
The invariant amplitude corresponding to quark level transition
$b\longrightarrow s \gamma\gamma$ (with an incoming $b$ line and an
outgoing
$s$ line and the two photons being emitted by the quark loop) is
\bea
{\cal{M}}_{b\rightarrow s} &=& \Bigg[\frac{16\sqrt{2}\alpha
G_FV_{ts}^*V_{tb}}{9\pi}
           \Bigg]\bar{u}(p_s)\Bigg\{\sum_q
A_qJ(m_q^2)\gamma^{\rho}P_L 
R_{\mu\nu\rho}\\ \nonumber
&+& \iota B(m_sK(m_s^2)P_L + m_bK(m_b^2)P_R) T_{\mu\nu}\\ \nonumber
&+& C(-m_sL(m_s^2)P_L + m_bL(m_b^2)P_R)\epsilon_{\mu\nu\alpha\beta}
k_1^{\alpha}k_2^{\beta}\Bigg\}u(p_b){\epsilon^{\mu}}^*(k_1){\epsilon^{\nu}}^*(k_2),
\eea
where
\[
R_{\mu\nu\rho} =
{k_1}_{\nu}\epsilon_{\mu\rho\sigma\lambda}k_1^{\sigma}
k_2^{\lambda} -
{k_2}_{\mu}\epsilon_{\nu\rho\sigma\lambda}k_1^{\sigma}
k_2^{\lambda} + (k_1.k_2)\epsilon_{\mu\nu\rho\sigma}(k_2-k_1)^{\sigma},
\]
\[
T_{\mu\nu} = {k_2}_{\mu}{k_1}_{\nu} - (k_1.k_2) g_{\mu\nu},
\]
\[
A_u = 3(C_3-C_5) + (C_4-C_6),
\]
\be
A_d = \frac{1}{4}[3(C_3-C_5) + (C_4-C_6)],
\ee
\[
A_c = 3(C_1+C_3-C_5) + (C_2+C_4-C_6),
\]
\[
A_s = A_b = \frac{1}{4}[3(C_4+C_3-C_5) + (C_3+C_4-C_6)],
\]
and
\[
B = C = -\frac{1}{4}(3C_6+C_5).
\]

To get to ${\cal{M}}(B\rightarrow K\gamma\gamma)$ from the quark
level
amplitude $({\cal{M}}(b\rightarrow s\gamma\gamma))$, we should
replace
$\langle s\vert\Gamma\vert b\rangle$ by $\langle K\vert\Gamma\vert
B\rangle$
for any Dirac bilinear $\Gamma$.
Recall that $\langle s\vert\bar{s}\gamma^{\rho}b\vert b\rangle$
gives a 
factor $\bar{u}_{s}\gamma^{\rho}u_{b}$ and similarly for 
$\bar{s}\gamma^{\rho}\gamma^5b$, $\bar{s}\gamma^5b$, and $\bar{s}b$, and
also that
$\langle K\vert\Gamma\vert B\rangle$ is zero for $\Gamma=\gamma^5,
~\gamma^{\rho}\gamma^5$. In this way we get
\bea
{\cal{M}}_{\rm{irred}}(B\rightarrow K\gamma\gamma) &=& 
\Bigg(\frac{16\sqrt{2}\alpha G_FV_{ts}^*V_{tb}}{9\pi}
           \Bigg)\Bigg[\frac{1}{2}\langle
K\vert\bar{s}\gamma^{\rho}b
\vert B\rangle\sum_q A_qJ(m_q^2)R_{\mu\nu\rho} \nonumber \\
&+&\frac{1}{2}\langle K\vert \bar{s}b\vert B\rangle\Bigg\{
\iota B(m_sK(m_s^2) + m_bK(m_b^2)) T_{\mu\nu}\\ \nonumber
&+& C(-m_sL(m_s^2) + m_bL(m_b^2))\epsilon_{\mu\nu\alpha\beta}
k_1^{\alpha}k_2^{\beta}\Bigg\}\Bigg]{\epsilon^{\mu}}^*(k_1){\epsilon^{\nu}}^*(k_2)
\eea
In the above expressions we introduced the functions
\[
J(m^2) = I_{11}(m^2), \hskip 1cm K(m^2) = 4I_{11}(m^2) - I_{00}(m^2),
\hskip 1cm     L(m^2) = I_{00}(m^2),
\] where
\be
I_{pq}(m^2) = \int_0^1dx\int_0^{1-x}dy \frac{x^py^q}{m^2 -
2(k_1.k_2)xy - \iota
\epsilon}
\ee
 We use the following parametrization for the matrix elements of the 
 quark vector current:
\bea
\langle K\vert\bar{s}\gamma_{\mu}b \vert B\rangle &=&
\Bigg((p_B+p_K)_{\mu}
- \frac{m_B^2-m_K^2}{q^2}q_{\mu}\Bigg)F_1^{BK}(q^2) \\ \nonumber
&+& \Bigg(\frac{m_B^2-m_K^2}{q^2}\Bigg)q_{\mu}F_0^{BK}(q^2),
\eea
with $q=p_b-p_K=k_1+k_2$. It then follows that
\bea
\langle K\vert\bar{s}b \vert B\rangle &=& (m_b-m_s)^{-1}
\Bigg[q_{\mu}\langle K\vert\bar{s}\gamma^{\mu}b \vert
 B\rangle \Bigg]\\ \nonumber
&=& (m_b-m_s)^{-1} (m_B^2-m_K^2)F_0^{BK}(q^2).
\eea
Also observe that
\bea
q^{\rho}R_{\mu\nu\rho} &=& (k_1+k_2)^{\rho}R_{\mu\nu\rho}\\
\nonumber
&=& (k_1.k_2)\epsilon_{\mu\nu\rho\sigma}(k_1^{\rho}k_2^{\sigma} - 
k_2^{\rho}k_1^{\sigma}).
\eea
 For the form factors appearing
above, we use the explicit  
numerical dependence of $F(q^2)$ on $q^2$ given by Cheng \emph{et al}
\cite{cheng}.
\end{section}
\begin{section}{Resonance contributions}
\subsection{The $\eta_c$  resonance}
 The $\eta_c$
contribution
to the deacy process comes via the t-channel decay $B\rightarrow
K \eta_c$,
 with then $\eta_c$ decaying into two photons. Since this has to be
added to other
contributions, we need to be careful about the sign of the
term.\par
The $T$-matrix element for this process can be written as
\be
\imath\langle K\gamma\gamma\vert T \vert B\rangle =
       \langle K\eta_c\vert -\imath{\cal{H}} \vert
B\rangle\frac{\imath}
{q^2-m_{\eta_c}^2+
\imath m_{\eta_c}\Gamma^{\eta_{c}}_{\rm{total}}}
\langle\gamma\gamma\vert-\imath{\cal{H}}
\vert\eta_c\rangle .     \ee
 Noting that in the lowest order, for any states $X,~Y$
\[
 \langle X  \vert T \vert Y \rangle = -\langle X\vert
{\cal{H}}\vert Y\rangle  ,
\]
we get
\be
\langle K\gamma\gamma\vert T \vert B\rangle =
-
\frac{\langle K\eta_c\vert T\vert B\rangle
\langle\gamma\gamma\vert T\vert\eta_c\rangle}{q^2-m_{\eta_c}^2 +
\imath m_{\eta_c}\Gamma^{\eta_{c}}_{\rm{total}}},
\ee
The amplitude $\langle\gamma\gamma\vert T\vert\eta_{c}\rangle$ is
parametrized
as \cite{reina}
\be
\langle\gamma\gamma\vert T\vert\eta_c\rangle = 2\imath B_{\eta_c}
\epsilon^{\mu\nu\alpha\beta}{\epsilon_1}_{\mu}^*{\epsilon_2}_{\nu}
^*
{k_1}_{\alpha}{k_2}_{\beta}. \label{eq1}
\ee
We can determine $B_{\eta_{c}}$ from the $\eta_{c} \to \gamma \gamma $ 
decay rate:
\be
\Gamma (\eta_c\rightarrow\gamma\gamma) = \Bigg(\frac{1}{2}\Bigg)
\frac{1}{2m_{\eta}}\int
\frac{d^3k_1}{(2\pi)^32k_1^0}\frac{d^3k_2}{(2\pi)^32k_2^0}(2\pi)^2
\delta^4(k_{\eta}-k_1-k_2)\vert \langle\gamma\gamma\vert
T\vert\eta_c\rangle
\vert^2.
\ee
We  have
\[
\sum_{spins}\vert \langle\gamma\gamma\vert T\vert\eta_c\rangle\vert^2
= 
2\vert B_{\eta_{c}}\vert^2 m_{\eta_c}^2,
\]
and so
\[
\Gamma (\eta_{c}\rightarrow\gamma\gamma) = \frac{\vert B_{\eta}\vert^2
m_{\eta_c}^3}
{16\pi}.
\]
Next we need the $B \to K \eta_{c}$ amplitude
\[
\langle K\eta_c\vert T\vert B\rangle = -\langle K\eta_c\vert
{\cal{H}}_{\rm{eff}}
\vert B\rangle.
\]
The relevant piece of ${\cal{H}}_{\rm{eff}}$ for this matrix element is
\bea
{\cal{H}}_{\rm{eff}} &=& \frac{G_F}{\sqrt{2}}V_{cb}V_{cs}^*
\Bigg[C_2(\bar{c}b)_{V-A}(\bar{s}c)_{V-A} + C_1(\bar{c}_ib_j)_{V-A}
(\bar{s}_jc_i)_{V-A}\Bigg] \hskip 0.25cm \\ \nonumber
&=& -\frac{G_F}{\sqrt{2}}V_{cb}V_{cs}^*
\Bigg[C_1(\bar{c}c)_{V-A}(\bar{s}b)_{V-A} + C_2(\bar{c}_ic_j)_{V-A}
(\bar{s}_jb_i)_{V-A}\Bigg].
\eea
The second expression is obtained by Fierz transforming the first
one. Thus
\be
{\cal{H}}_{\rm{eff}} =
-\frac{G_F}{\sqrt{2}}V_{cb}V_{cs}^*(C_1+\frac{C_2}{3})
(\bar{c}c)_{V-A}(\bar{s}b)_{V-A}.
\ee
Between $\vert B\rangle$ and $\vert K\rangle$ only the V part of
$(\bar{s}b)_
{V-A}$ contributes whereas for the c-loop with two photons only the
A part of
$(\bar{c}c)_{V-A}$ is relevant. Hence,
\be
{\cal{H}}_{\rm{eff}} =
\frac{G_F}{\sqrt{2}}V_{cb}V_{cs}^*(C_1+\frac{C_2}{3})
(\bar{c}\gamma_{\mu}\gamma_5c)(\bar{s}\gamma^{\mu}b).
\ee
Using factorization we can write
\be
\langle K\eta_c\vert T\vert B\rangle =
\frac{G_F}{\sqrt{2}}V_{cb}V_{cs}^*
(C_1+\frac{C_2}{3})\langle K\vert\bar{s}\gamma^{\mu}b\vert B\rangle
\langle\eta_c\vert\bar{c}\gamma_{\mu}\gamma_5c\vert 0\rangle.
\ee
Define
\[
\langle 0\vert A_{\mu}^c\vert\eta_c\rangle \equiv \imath
f_{\eta_{c}}q_{\mu}
\hskip 1cm A_{\mu}^c = \bar{c}\gamma_{\mu}\gamma_5 c
\]
\[
q^{\mu}\langle K\vert\bar{s}\gamma_{\mu}b\vert B\rangle =
F_0(m_{\eta_c}^2)
(m_B^2-m_K^2)
\]
We thus get
\be
\langle K\eta_c\vert T\vert B\rangle =
-\imath\frac{G_F}{\sqrt{2}}V_{tb}V_{ts}^*
f_{\eta_c}F_0(m_{\eta_c}^2)(m_B^2-m_K^2)(C_1+\frac{C_2}{3})
\ee
where a dipole form of the form factor $F_0(m_{\eta_c}^2)$ is used.

Now we fix the relative sign between $B_{\eta_c}$ and $f_{\eta_c}$.
This can be 
easily done through the anomaly equation \cite{chengli}
\bea
\langle\gamma\gamma\vert T\vert\eta_c\rangle &=&
{\epsilon_1^*}^{\mu}(k_1)
{\epsilon_2^*}^{\nu}(k_2)\Gamma_{\mu\nu}\\ \nonumber
&=& {\epsilon_1^*}^{\mu}(k_1){\epsilon_2^*}^{\nu}(k_2)\frac{\imath
e^2D}{2\pi^2
f_{\eta_c}}\epsilon_{\mu\nu\alpha\beta}k_1^{\alpha}k_2^{\beta}
\eea
Comparing with our definition of $B_{\eta_c}$ in eq.(\ref{eq1}), we 
identify 
\[
B_{\eta_c} = \frac{e^2D}{4\pi^2f_{\eta_c}},
\]
and see that the relative sign is positive.
However, we emphasise that we do not use this theoretical result to 
determine the numerical value of either $B_{\eta_c}$ or $f_{\eta_c}$, 
but merely use it to fix the relative sign.

 The total contribution due to $\eta_c$ resonances
is thus \bea {\cal{M}}_{\eta_c} &=&
2B_{\eta_c}f_{\eta_c}\frac{G_F}{\sqrt{2}}
V_{tb}V_{ts}^*(C_1+\frac{C_2}{3})F_0(m_{\eta_c}^2)(m_B^2-m_K^2) 
\label{eq-eta-c}\\
\nonumber && {\epsilon_1^*}^{\mu}(k_1){\epsilon_2^*}^{\nu}(k_2)
\epsilon_{\mu\nu\alpha\beta}k_1^{\alpha}k_2^{\beta}
\frac{1}{q^2-m_{\eta_c}^2+ \imath
m_{\eta_c}\Gamma_{\rm{total}}^{\eta_c}}. \eea
\subsection{The $\eta$ contribution}

Analogous to $\eta_c$, the $\eta$-resonance will
also contribute to the decay amplitude because of its coupling to 
$c\bar{c}$ channel. This contribution,
$\mathcal{M}_{\eta}$,has exactly the same form as eq.(\ref{eq-eta-c})
with the parameters $B_{\eta_c}$, $f_{\eta_c}$, $F_0(m_{\eta_c}^2)$,
$m_{\eta_c}$ and $\Gamma_{\rm{total}}^{\eta_c}$ being replaced by their
$\eta$-counterparts.  However in this case we cannot define the 
relative sign of the $\eta$ and any of the other components of the 
amplitude.
\subsection{The $\eta^{\prime}$ contribution}

Like $\eta_c$ and $\eta$, the $\eta^{\prime}$ can also contribute via
the effective Hamiltonian, eq.(15).  However, unlike $\eta_c$ and
$\eta$, the $\eta^{\prime}$ has a very strong coupling to a two gluon
state.  Theoretical models for $\eta^{\prime}$ production in B-decay
via $gg$-states exist both when the gluons arise from a basic
$b\rightarrow sgg$ process \cite{ali} and also when the spectator
quark in B-meson emits a gluon to combine with a basic $b\rightarrow
sg$ process \cite{ahmady1,ahmady2}.  These estimates have considerable
theoretical uncertainities.  Fortunately, experimental data on
$B\rightarrow K\eta^{\prime}$ exists, which is sufficient for our
purpose since the invariant mass of the two photons will be strongly
peaked at $m_{\eta^{\prime}}^2$.  Calling the coupling of
$B^i\rightarrow K^i\eta^{\prime}$ ($i=+,0$) as
$F(B^iK^i\eta^{\prime})$, the $\eta^{\prime}$ contribution to our
amplitude can be written as \be {\mathcal{M}}_{\eta^{\prime}} = 2
B_{\eta^{\prime}}F(B^iK^i\eta^{\prime})
{\epsilon_1^*}^{\mu}(k_1){\epsilon_2^*}^{\nu}(k_2)
\epsilon_{\mu\nu\alpha\beta}k_1^{\alpha}k_2^{\beta}
\frac{1}{q^2-m_{\eta^{\prime}}^2+ \imath
m_{\eta^{\prime}}\Gamma_{\rm{total}}^{\eta^{\prime}}} \ee where
$B_{\eta^{\prime}}$ is defined as in eq.(12) with
$\eta_c\rightarrow\eta^{\prime}$.  Also, $F(B^iK^i\eta^{\prime})$ is
related to the decay rate $\Gamma(B^i\rightarrow K^i\eta^{\prime})$
as: \be \Gamma(B^i\rightarrow K^i\eta^{\prime}) = \frac{1}{16\pi m_B}
\vert F(B^iK^i\eta^{\prime})\vert^2
\lambda^{\frac{1}{2}}(1,\frac{m_K^2}{m_B^2},\frac{m_{\eta^{\prime}}^2}{m_B^2})
\ee

Once again the relative sign of the contribution is unknown.

\subsection{$K^*$ resonance contribution}
In this case we have the following situation:
\[
B^i(p_B) \longrightarrow {K^*}^i + \gamma (k_1), \hskip 1.5cm i~=~+,~0
\]
followed by
\[
{K^*}^i \longrightarrow K^i + \gamma(k_2),
\]
and the process with $k_1 \leftrightarrow k_2$.

The effective Hamiltonian for the first vertex is
\be
{\cal{H}}_{\rm{eff}} =
4\frac{G_F}{\sqrt{2}}V_{cb}V_{cs}^*C_7\Bigg(\frac{em_b}
{16\pi^2}\Bigg)\bar{s}_L\sigma_{\mu\nu}b_R F^{\mu\nu}
\ee
and the $T$-matrix element $T^i$ for the process ${K^*}^i
\rightarrow K^i +  
\gamma$ is
\be
T^i =
g^i_{K^*K\gamma}\epsilon_{\delta}(p_B-k_1)\epsilon^*_{\beta}(k_2)
{k_2}_{\alpha}(p_B-k_1)^{\lambda}\epsilon_{\alpha\beta\lambda\delta}
\ee
with $\vert g^i_{K^*K\gamma}\vert$ to be determined from the decay 
rate
of 
${K^*}^i \rightarrow K^i +  \gamma$. Summing over the final spins
and averaging
over the initial spins we get
\be
\frac{1}{3}\sum\vert T^i\vert^2 = \frac{1}{3}\vert g^i\vert^2 
\frac{(m_{K^*}^2-m_{K}^2)^2}{2},
\ee
so $|g_{i}|$ is determined from
\be
\Gamma^i = \frac{\vert
g^i\vert^2}{96\pi}\Bigg(\frac{m_{K^*}^2-m_{K}^2}
{m_{K^*}}\Bigg)^3.
\ee

The matrix element of 
\[
O_7 = \Bigg(\frac{em_b}{16\pi^2}\Bigg)\bar{s}_L\sigma_{\mu\nu}b_R
F^{\mu\nu}
\]
is
\bea
\langle {K^*}^i(q)\vert O_7\vert B^i(p+q)\rangle &=& -\imath
\Bigg(\frac{em_b}{32\pi^2}\Bigg)F^i[\epsilon^{\alpha}(p)p^{\beta} -
\epsilon^{\beta}(p)p^{\alpha}] \\ \nonumber
&&\epsilon^{\sigma}(q)~[\imath\epsilon_{\alpha\beta\sigma\tau}q^{\tau} -
2g_{\sigma\alpha}q_{\beta}]
\eea
with $F^i$ to be determined from the radiative decay $B^i
\rightarrow 
{K^*}^i \gamma$. With these definitions the complete $T$-matrix
element
can be written as
\bea
\langle K\gamma\gamma\vert T \vert B\rangle &=& {\cal{M}}_{K^*} \\ \nonumber
&=& \Bigg[T^{\mu\nu}(k_1,k_2) + 
(\mu \leftrightarrow \nu,~k_1 \leftrightarrow k_2) \Bigg]
{\epsilon^*}_{\mu}(k_1){\epsilon^*}_{\nu}(k_2)
\eea
with
\bea
T^{\mu\nu}(k_1,k_2) &=& -\Bigg(\frac{em_bg^iF^i}{16\pi^2}\Bigg)
4\frac{G_F}{\sqrt{2}}V_{cb}V_{cs}^*C_7
\epsilon^{\alpha\nu\gamma\delta}
{k_2}_{\alpha}(p_B-k_1)_{\gamma}{k_1}_{\beta^{\prime}} \\ \nonumber
&&\Bigg[\frac{\Bigg(g_{\delta\sigma^{\prime}}-\frac{(p_B-k_1)_
{\delta}(p_B-k_1)_{\sigma^{\prime}}}{m_{K^*}^2}\Bigg)}
{(p_B-k_1)^2-m_{K^*}^2+\imath m_{K^*}\Gamma^{K^*}_{\rm{total}}}\Bigg]\\
\nonumber
&&
\Bigg[\imath\epsilon^{\mu\beta^{\prime}\sigma^{\prime}\tau^{\prime}}
(p_B-k_1)_{\tau^{\prime}} -
(g^{\mu\sigma^{\prime}}(p_B-k_1)^{\beta^{\prime}}
-g^{\beta^{\prime}\sigma^{\prime}}(p_B-k_1)^{\mu})\Bigg].
\eea
In the last expression we have used the antisymmetry of
$F^{\mu\nu}$ to
write the terms in this particular form. 

Also note that the sign of this 
contribution can not be unambiguously fixed as was donefor the 
$\eta_{c}$ contribution.

\end{section}

\begin{section}{Results}
The total contribution to the process $B \rightarrow K
\gamma\gamma$ is 
\be
{\cal{M}}_{tot} = {\cal{M}}_{irred} +  
{\cal{M}}_{\eta_c} + \xi_{K^{*}} {\cal{M}}_{K^*} +\xi_{\eta} {\cal{M}}_{\eta} +
\xi_{\eta^{\prime}}{\cal{M}}_{\eta^{\prime}} 
\ee
where $\xi_{\alpha}=\pm 1$ are sign parameters, introduced because 
the signs of these terms are unknown.
The total decay rate is given by 
\be
\frac{d\Gamma}{d\sqrt{s_{\gamma\gamma}}} = \Bigg(\frac{1}{512m_B\pi^3}\Bigg)
\sqrt{s_{\gamma\gamma}}
\Bigg[\Bigg(1 - \frac{s_{\gamma\gamma}}{m_B^2} + \frac{m_K^2}{m_B^2}\Bigg)^2
 - \frac{4m_K^2}{m_B^2}\Bigg]^{\frac{1}{2}}\int_0^{\pi}d\theta~\sin\theta
 \vert{\mathcal{M}}_{tot}\vert^2
\ee
where $\sqrt{s_{\gamma\gamma}}$ is the C.M.
energy of the two photons
while $\theta$ is the angle which the decaying B-meson makes with one of the 
two photons in the $\gamma\gamma$ C.M. frame. The numerical values
of various parameters used in our calculation are listed in Appendix.

Figure 1 shows the spectrum of our results given as a function
of the invariant mass of the two photons, for the case of neutral $B$ 
meson decay. The figure shows the
expected resonance peaks in the $\gamma \gamma$ spectrum at the 
positions of the $\eta$, $\eta^{\prime}$, and $\eta_{c}$, and the $K^*$ 
resonance contribution is spread out across this projection of the 
Dalitz plot. We have chosen the sign parameters $\xi_{\alpha} = +1$ in 
all cases in the graph. Figure 2 contrasts the irreducible and the total
contribution to the branching ratio as a function of momentum transferred
squared over the whole kinematic range.

The various contributions to the branching ratio are given in Table 
1, with the maximum branching ratio ($\xi_{\alpha} = +1$ for all 
$\alpha$), and the minimum branching ratio ($\xi_{\alpha} = -1$ for 
all $\alpha$).  We see that we predict the branching ratio in the 
range 
\be
1.439 \times 10^{-6} \le \rm{Br}(B \to K \gamma \gamma) \le 1.485
\times 10^{-6}
\ee
and that the largest single contribution to the branching ratio comes 
from the $\eta'$ resonance term.  In fact the
resonance terms (neglecting interference) are far far greater than
the ireducible term.
\begin{table}[ht]
    \caption{Contributions to the $B^{0}\to K^{0}\gamma\gamma$ 
    branching ratio --- the total branching ratio and with a cut on 
    $\sqrt{s_{\gamma\gamma}}$}
    \begin{center}

    \begin{tabular}{|c|c|c|}
	\hline
	Contribution & Branching ratio & With cut \\
	& $\times 10^{-7}$ & $\times 10^{-7}$ \\
	\hline
	Resonance & & \\
	$ \eta_c $	& 3.86	& 1.94\\ 
$\eta$	& 0.004 & 0	\\ 
$\eta'$	& 9.69	& 0.005\\ 
$K^{*}$	& 1.07	 & 0.56\\ 
& & \\
Irreducible &	0.0195	& 0.0193\\ 
& & \\
Interference & &\\
$\eta_c - I$	&	-0.025 & 0.076\\ 
$\eta' - I$	& $\mp$	0.13 & 0\\ 
$\eta_c - K^{*}$	& $\mp$	0.048 & $\mp$ 0.099\\ 
$\eta' - K^{*}$	& $\pm$	0.054 & $\pm$ 0.018\\ 
$ K^{*} - I$	& $\mp 1.5\times 10^{-4}$&  $\mp 1.43\times 10^{-4}$\\
\hline
Maximum BR & 14.85 & 2.72 \\
Minimum BR & 14.39 & 2.48 \\
\hline
\end{tabular}
\end{center}
\end{table}

In an attempt to enhance the relative contribution of the irreducible 
term we have calculated the partial branching ratio with a cut on 
$\sqrt{s_{\gamma\gamma}} > m_{\eta_{c}} + 2 \Gamma_{\eta_{c}} \approx 
3.02~ \rm{GeV}$.  In this partial branching ratio the $\eta_c$ term 
does dominate and is about four the $K^*$ contribution.  We could try to
further reduce the non-irreducible background by introducing a 
further cut in the Dalitz plot to stay above the $K^*$ peak in 
$\sqrt{s_{K\gamma}}$, but because of the already small branching ratio, 
we do not suggest this additional cut. \\ 

For completeness, we also quote the analogous 
individual contributions to the branching
ratio for the
corresponding charged decay mode ($B^+ \rightarrow K^+ \gamma \gamma$) in
Table 2.

\begin{table}[ht]
    \caption{Contributions to the $B^{+}\to K^{+}\gamma\gamma$ 
    branching ratio --- the total branching ratio and with a cut on 
    $\sqrt{s_{\gamma\gamma}}$}
    \begin{center}

    \begin{tabular}{|c|c|c|}
	\hline
	Contribution & Branching ratio & With cut \\
	& $\times 10^{-7}$ & $\times 10^{-7}$ \\
	\hline
	Resonance & & \\
	$ \eta_c $	& 3.85	& 1.94\\ 
$\eta$	& 0.0037 & 0	\\ 
$\eta'$	& 12.61	& 0.005\\ 
$K^{*}$	& 0.9	 & 0.46\\ 
& & \\
Irreducible &	0.0196	& 0.0193\\ 
& & \\
Interference & &\\
$\eta_c - I$	&	-0.025 & 0.076\\ 
$\eta' - I$	& $\mp$ 0.13 & 0\\ 
$\eta_c - K^{*}$	& $\mp$	0.043 & $\mp$ 0.082\\ 
$\eta' - K^{*}$	& $\pm$	0.067 & $\pm$ 0.018\\ 
$ K^{*} - I$	& $\mp 1.6 \times 10^{-4}$ &  $\mp 1.47 \times 10^{-4} $ \\
\hline
Maximum BR & 17.6 & 3.61 \\
Minimum BR & 17.12 & 3.41 \\
\hline
\end{tabular}
\end{center}
\end{table}

Comparing the the two branching ratios for the neutral and charged deacy
modes, one finds that although the $\eta^{\prime}$ contribution is larger
in the case of the charged mode, the overall branching fractions are 
almost identical, whether one considers the complete branching ratio, 
or just the part above the cut.
 This is because of the fact that the larger branching
fraction of the charged mode $B^+ \rightarrow K^{*+} \gamma$ gets compensated
by the smaller value of the branching ratio for the mode
$K^{*+} \rightarrow K^+\gamma$. \\

From the tables above, it is clear that above the imposed cut,
the $\eta$ and $\eta^{\prime}$ do not contribute significantly. 
But the same is not found to
be true for $\eta_c$. In particular, the interference terms are practically
unimportant, with or without the cut. This is essentially due to small
contribution from the irreducible sector while for interference between two
resonance contributions, the fact that the cut is practically far from the
resonances results in a negligible contribution.
 
The value obtained for $Br(B\rightarrow K\gamma\gamma)$, is much much higher
than the corresponding branching ratio for the inclusive
process $B\rightarrow X_s\gamma\gamma$, where the total branching ratio
is quoted to be $(1.7~-~3.37)\times 10^{-7}$ \cite{reina}.  Even the 
exclusive branching ratio with much of the long distance resonance 
effects eliminated by the cut in $\sqrt{s_{\gamma\gamma}}$ we have 
suggested, could exceed these values (or atleast be of similar size).
However, it is known \cite{falk} that the inclusive rate calculation based on 
quark transition picture is not valid when $M_X^2$ becomes low, which is true
in our case.  Thus we should not be too surprised by our larger 
result for just one part of the inclusive branching ratio.

 It is worth mentioning at this point
that if instead of using the explicit numerical dependence of $B\rightarrow K$
form factors, following
Chen etal\cite{chen} we assume $F_0^{BK}(0) = F_1^{BK}(0) = 0.341$
and
$F_0^{BK}(q^2)=\frac{F_0^{BK}(0)}{(1-\frac{q^2}{M_{pole}^2})^2}$
with $M_{pole}=6.65$ GeV \cite{fusaoka}, we get an even higher value
for the branching ratio.

Observation of this decay experimentally
thus is expected to  provide an interesting test of the underlying 
theory leading to this value.  In particular the branching ratio 
above a cut which removes the $\eta, \eta' $ and $\eta_{c}$ 
contributions will, ideally speaking,
 be a good test of the magnitude of the irreducible 
and will thus be sensitive to the effects of high mass intermediate 
states, even to physics beyond the standard model.

\end{section}

\section*{Acknowledgements} 
NM would like to thank the University Grants Commission,
India, for a fellowship. SRC would like to acknowledge support from
DST, Government of India, under the SERC scheme.

This work was supported in part by the Australian Research Council.

We would like to thank Prof. G.Hiller for a communication pointing out 
discrepancies between their results and our earlier results for
the irreducible contribution.

\begin{section}*{Appendix}
We give the input parameters used in the numerical calculations.\\
\begin{table}[ht]
\begin{center}
\begin{tabular}{|l|l|l|l|l|l|l|l|}\hline
$C_1$&$C_2$&$C_3$&$C_4$&$C_5$&$C_6$&$C_7$&$C_8$\\ \hline
-0.222&1.09&0.010&-0.023&0.007&-0.028&-0.301&-0.144 \\ \hline
\end{tabular}
\caption{The approximate values of $C_i's$ at $\mu=m_b$}
\end{center}
\end{table}
\[
m_b=4.8~GeV \hskip 1cm m_c=1.5~GeV \hskip 1cm m_t=175~GeV
\]
\[
m_s=0.15~GeV \hskip 1cm m_u =m_d =0
\]
\[
m_{B^0}=5.2792~GeV \hskip 1cm \Gamma^{B^0}_{\rm{total}}=4.22\times 10^{-13}~GeV
\]
\[
m_{B^+}=5.2789~GeV \hskip 1cm \Gamma^{B^+}_{\rm{total}}=4.21\times 10^{-13}~GeV
\]
\[
m_{K^0}=0.497672~GeV \hskip 1.5cm  m_{K^+}=0.493677~GeV
\]
\[
m_{\eta_c}=3~GeV \hskip 1.5cm B_{\eta_c}=2.74\times 10^{-3}~GeV^{-1}
\]
\[
\Gamma_{\rm{total}}^{\eta_c}=1.3\times 10^{-2}~GeV \hskip 1.5cm 
f_{\eta_c}=0.35~GeV
\]
\[
m_{\eta}=0.547~GeV \hskip 1.5cm \Gamma_{\rm{total}}^{\eta}=1.18\times 10^{-6}~GeV
\]
\[
B_{\eta}=13.254\times 10^{-3}~GeV^{-1}
\hskip 1.5cm  f_{\eta}=-2.4\times 10^{-3}~GeV
\]
\[
m_{\eta^{\prime}}=0.95778~GeV \hskip 1.5cm
\Gamma_{\rm{total}}^{\eta^{\prime}}=0.203\times 10^{-3}~GeV
\]
\[ 
Br(B^0\to K^0\eta^{\prime})=4.7\times 10^{-5} \hskip 1.5cm
Br(B^+\to K^+\eta^{\prime})=6.5\times 10^{-5}
\]
\[
m_{K^{*0}}=0.896~GeV \hskip 0.5cm g_{K^{*0}K\gamma}=0.384916~GeV^{-1} 
\hskip 0.5cm 
\Gamma_{\rm{total}}^{K^{*0}}=50.5~MeV
\hskip 0.5cm F^i=0.52-0.72
\]
\[
m_{K^{*+}}=0.8916~GeV \hskip 0.5cm g_{K^{*+}K\gamma}=0.355~GeV^{-1} 
\hskip 0.5cm 
\Gamma_{\rm{total}}^{K^{*+}}=50.8~MeV
\]
We follow the Wolfenstein parametrization of the CKM matrix with
\[
A=0.8 \hskip 1cm \lambda=0.22 \hskip 1cm \eta=0.34
\]
\[
V_{tb}\sim 1 \hskip 1cm V_{ts}=-A\lambda^2
\]
\[
V_{cb}=A\lambda^2 \hskip 1cm V_{cs}=1-\frac{\lambda^2}{2}
\]
\end{section}
\begin{figure}[ht]
\vspace*{-1cm}
\centerline{
\epsfxsize=12.5cm\epsfysize=10.5cm
                     \epsfbox{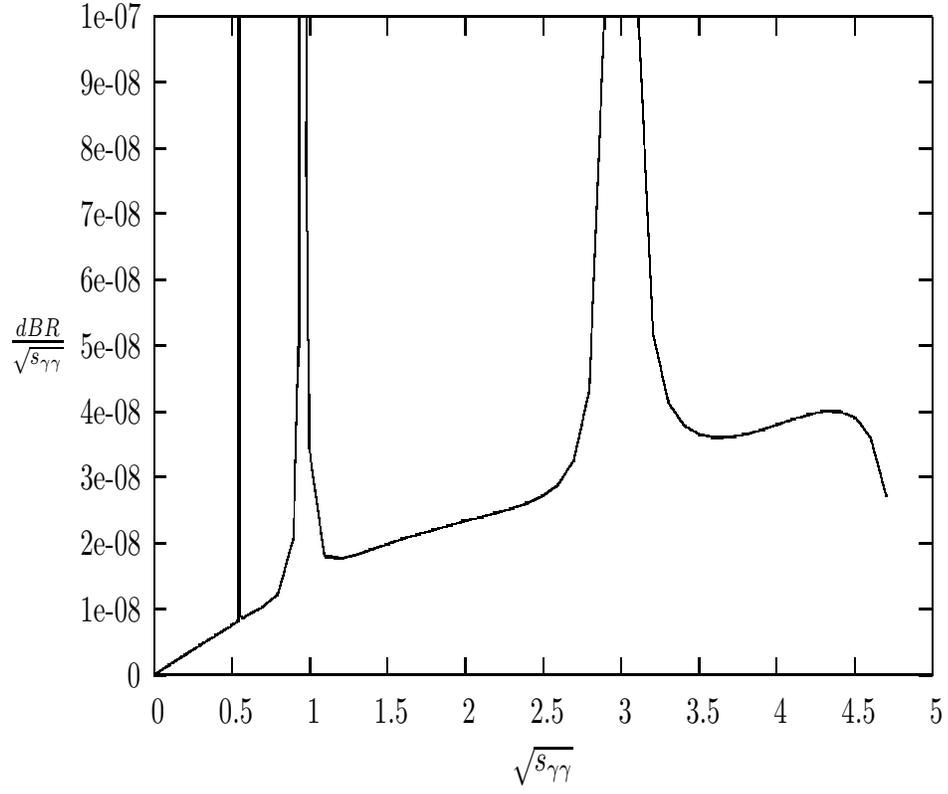}}
\vskip 1.5cm
\centerline{
\epsfxsize=12.5cm\epsfysize=10.5cm
                     \epsfbox{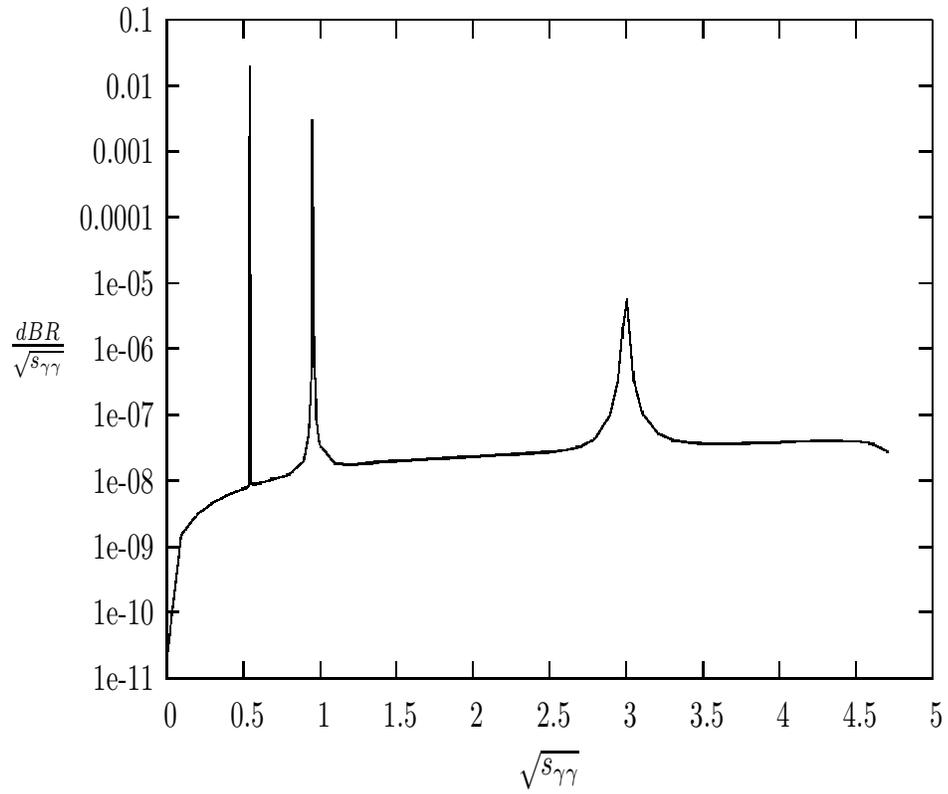}}
\caption{\em Our result for  the spectrum of $B^0\rightarrow K^0 \gamma\gamma$
(above). The same results plotted with logscale on y-axis (below).
 The parameters used are
listed in the appendix. 
	}
\label{fig:fig1}
\end{figure}
\begin{figure}[ht]
\vspace*{-1cm}
\centerline{
\epsfxsize=12.5cm\epsfysize=10.5cm
                     \epsfbox{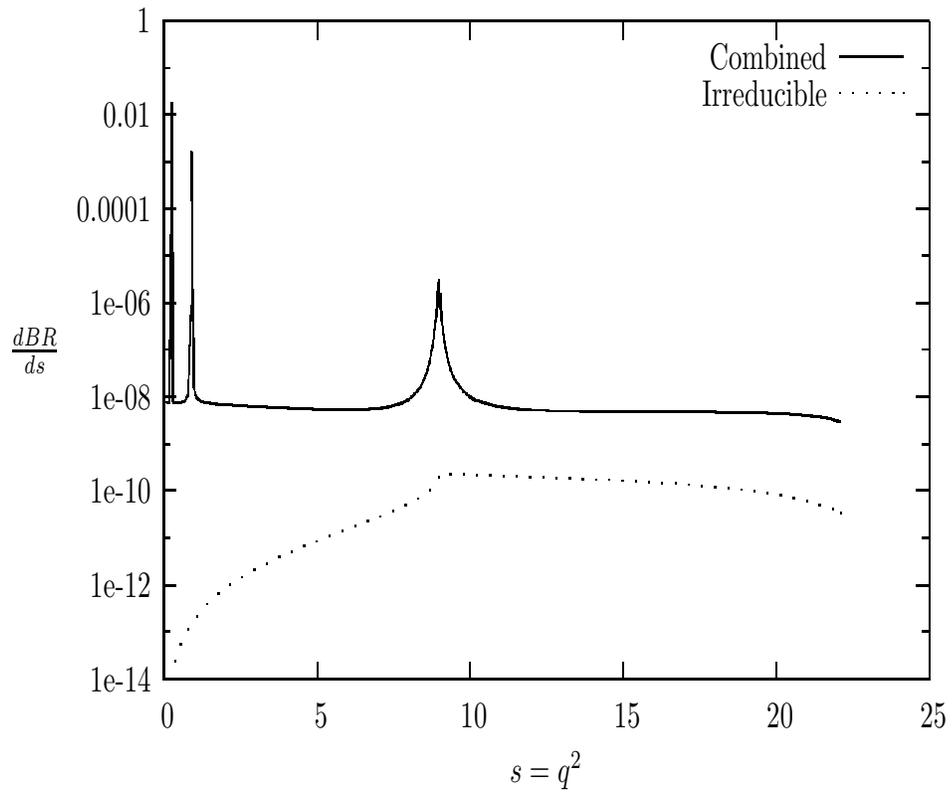}}
\caption{\em The spectrum of $B^0\rightarrow K^0 \gamma\gamma$ with and 
without resonances. We plot the differential BR as a function of momentum
transferred squared over the whole kinematic regime. 
	}
\label{fig:fig2}
\end{figure}

\end{document}